\documentclass{PoS}

\usepackage{amsmath} 

\title{Dual Methods for Lattice Field Theories at Finite Density}

\ShortTitle{Dual Methods for Lattice Field Theories at Finite Density}

\author{\speaker{Thomas Kloiber}, Christof Gattringer\\
        Institut f\"ur Physik, Karl-Franzens-Universit\"at, 8010 Graz, Austria \\
        \email{thomas.kloiber@uni-graz.at} \\ \email{christof.gattringer@uni-graz.at}}

\abstract{We present a dual representation of the partition function of the charged scalar field in which the complex action problem at non-zero chemical potential is absent. In this dual representation Monte Carlo simulations are possible and we show some physical results obtained with this approach. Furthermore we present a technique to study 2-point functions at finite density. Results for the lattice correlators at various chemical potentials are shown and discussed.}

\FullConference{31st International Symposium on Lattice Field Theory - LATTICE 2013\\
		July 29 - August 3, 2013\\
		Mainz, Germany}

\begin{document}

\section{Introduction}

In elementary particle physics, there are many interesting phenomena associated with non-zero or even very high (anti-)particle densities, like processes in the interior of neutron stars or nothing less than the genesis of the universe. It is believed, that matter undergoes certain phase transitions in the temperature and chemical potential plane.
From a QCD point of view, we essentially expect a confined phase, where hadrons are the fundamental building blocks, and a deconfined phase, where quarks and gluons form the so-called quark-gluon plasma (e.g., \cite{BMW09}).
However, details of this phase structure are unknown, except along the temperature axis, i.e., at zero chemical potential \cite{AEFKS06}.

The reason for this situation is a fundamental problem appearing at non-zero chemical potential, with no general solution \cite{TW05}, known as the sign problem. In lattice QCD, the problem can be traced back to the loss of \(\gamma_5\) hermiticity of the Dirac matrix for finite chemical potential \(\mu\), which in turn gives rise to a complex fermion determinant. This means, that standard importance sampling methods cannot be applied, because there the determinant enters the probability weight for a given configuration and for \(\mu \neq 0\) we lose the probabilistic interpretation. A very similar situations arises in other lattice field theories, where the action (and therefore the Boltzmann factor) becomes complex for non-zero chemical potential

There exist some approaches to tackle this problem which are, e.g., reviewed in \cite{DF09}. More recently it was realized, that the sign problem is basis dependent, i.e., there may exist some representations of the partition function (essentially one switches from one set of degrees of freedom to another one), where it is built up from real, non-negative contributions only and thus making a probabilistic interpretation feasible \cite{E07}. Henceforth we will refer to such a reformulation as a dual representation. Unfortunately, up to now it could only be shown for relatively simple theories (essentially all abelian in nature) that they admit a dual representation in which the sign problem is gone.

We have shown recently, that a charged scalar field with a quartic self-interaction can be treated in this way \cite{GK13} and addressed the problem of calculating \(n\)-point functions and generalized this approach \cite{GK13b}. Up to now, an extension of this formalism to more complex non-abelian models is still missing.

\section{Dual Representation of the Charged Scalar Field}
\label{sec_dual}

In the conventional representation the theory of a charged scalar field coupled to a chemical potential with a quartic self-interaction is defined by the Euclidean continuum action
\begin{equation}
S=\int \!\! d^4x ~ \Big[ |\partial_\nu \phi|^2  + (m^2-\mu^2)\,|\phi|^2+ \lambda\,|\phi|^4 +\mu\,(\phi^*\partial_4\phi-\partial_4\phi^*\phi)\Big] ~.
\end{equation}
Discretization leads to the lattice action
\begin{equation}
S = \sum_x \left(- \sum_{\nu=1}^4 \left(e^{\mu \; \! \delta_{\nu,4}} \, \phi_x\phi^*_{x+\hat{\nu}} +e^{ -\mu \; \! \delta_{\nu,4}} \, \phi^*_x\phi_{x+\hat{\nu}}\right)  + \kappa \, |\phi_x|^2 + \lambda \, |\phi_x|^4  ~\right)~,
\end{equation}
where the chemical potential was treated according to \cite{HK83}. The bare parameters of the theory are the rescaled mass \(\kappa \equiv m^2+8\), the quartic self-coupling \(\lambda\) and the chemical potential \(\mu\). The lattice coordinate \(x\) runs over the sites of a \(N_s^3\times N_t\) lattice, the directional index \(\nu\) over the four space-time directions and we implicitly assume periodic boundary conditions and a lattice constant set to unity (\(a\equiv1\)).

The complex action problem is manifest in this expression. Taking \(\mu=0\), we see that the first two terms are complex conjugate to each other, leading to a real term in the sum. A non-vanishing chemical potential spoils this property and we get complex contributions to the action. The Euclidean lattice partition function is given by
\begin{equation}
Z = \int \mathcal{D}[\phi] ~ e^{-S[\phi]} ~~~~~ \text{with} ~~~~~ \int \mathcal{D}[\phi] \equiv \prod_x \int_{\mathbb{C}} \frac{d\phi_x}{2\pi}~,
\end{equation}
where the arbitrary normalization of the measure was chosen for later convenience. The partition function itself is a real quantity, but we see, if the action is complex for some field configurations we have complex contributions to \(Z\) which eventually cancel but spoil a probabilistic interpretation of the individual contributions and thus numerical simulations.

In \cite{GK13} it was shown, that one can find a different expression for \(Z\) in which it is build up from real, non-negative contributions only. The basic idea is to write the exponential of the sum of the nearest neighbor terms as a product over all lattice sites and directions of exponentials of local terms only, where each local term depends on the field at a given lattice site and on its neighbor in a given direction. Each local exponential is then expanded in a power series and the summation variables are labeled by \(x\) and \(\nu\), i.e., \(l_{x,\nu}\). The exponentials with terms depending only on one lattice site and coming with an even power, i.e., the mass and self-interaction terms, remain unexpanded. After this step one can reorder products, sums and integrals and integrate out the original degree of freedom, i.e., the field \(\phi\). Doing this in terms of angular and radial parts, we face two types of integrals. The angular integration gives rise to Kronecker deltas (written \(\delta(n)\)) and therefore imposes constraints on the summation variables \(l_{x,\nu}\). The integrand of the radial integration is a polynomially growing factor times an exponential damping factor, which -- as long as we have a positive coupling lambda -- always converges and is sufficiently smooth to be calculated numerically. It gives rise to weight factors denoted by \(\mathcal{W}(n)\).

In its final form the partition function in dual representation is given by
\begin{equation}\label{eq_part_dual}
\begin{split}
Z &= \sum_{\{k,m\}} ~ \prod_{x,\nu} \frac{1}{(|k_{x,\nu}|+m_{x,\nu})!~m_{x,\nu}!} ~~~ \prod_x \delta \left( \sum_\nu [k_{x,\nu}-k_{x-\hat{\nu},\nu}]\right) \\
& \times ~\prod_x e^{\mu k_{x,4}} ~ \mathcal{W}\left( \sum_\nu \Big[ |k_{x,\nu}| + |k_{x-\hat{\nu},\nu}| +2(m_{x,\nu}+m_{x-\hat{\nu},\nu}) \Big] \right)~,
\end{split}
\end{equation}
with two types of dual variables -- constrained \(k_{x,\nu} \in \mathbb{Z}\) and unconstrained \(m_{x,\nu} \in \mathbb{N}\) -- interpreted as new degrees of freedom, i.e., integer-valued fields living on the links of the lattice instead of complex fields sitting on the sites. In this form one finds, that each term in the sum over all configurations (\(\sum_{\{k,m\}}\)) is real and non-negative and thus gives a proper probability weight for a given configuration of the variables \(\{k,m\}\), which can be used in a Monte Carlo simulation.

Due to the appearance of Kronecker deltas in the ensemble (Eq. \ref{eq_part_dual}), numerical sampling based on a standard local update is bound to fail. Instead one can interpret the constraints as local conservation of \(k\)-flux, corresponding to closed loops as admissible configurations. Using the Prokof'ev-Svistunov worm algorithm \cite{PS01} such a system can be updated efficiently. The main idea is to start at a random lattice site, add links to a neighboring site, accept the new link according to the probability weight obtained from the partition function and respect the constraint. This step is repeated until one reaches the starting point again and one ends up with a new valid closed loop configuration.

Within this setting we can study any bulk observable. We only have to take the partial derivatives of the dual partition function with respect to the parameters of the theory. This is illustrated here by the particle number density, which is given by
\begin{equation}
n ~ \equiv ~\frac{T}{V} \frac{\partial \ln Z}{\partial \mu} ~ = ~ \frac{1}{N_s^3 ~ N_t} \frac{\partial \ln Z}{\partial \mu} ~ = ~ \frac{1}{N_s^3 ~ N_t} \left\langle \sum_x k_{x,4} \right\rangle~. 
\end{equation}
On the right hand side we find a particular simple expression in terms of dual variables. The particle number density is proportional to the (dual) expectation value of the sum of \(k\)-variables in time direction. An important observation here is that, as long as the \(k\)-variables close trivially (i.e., not around the compactified time direction), the sum in the expectation value is identically zero. The particle number density therefore can be interpreted as as the winding number of \(k\)-flux around the 4-direction, i.e., the compactified time.

\begin{figure}[h] 
\centering
\includegraphics[width=10cm,type=pdf,ext=.pdf,read=.pdf]{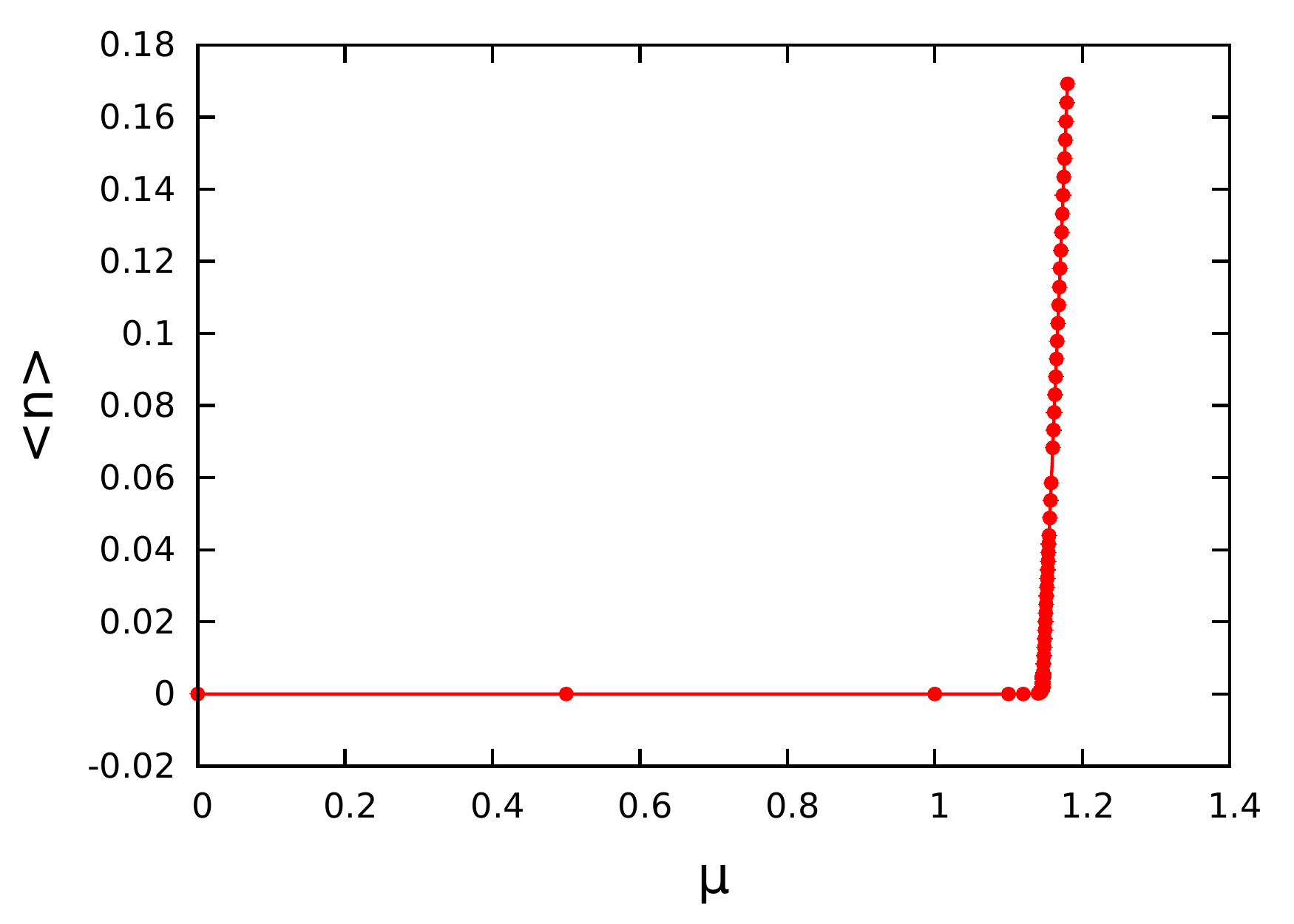}
\caption{Particle number density versus chemical potential on a \(N_s=20\), \(N_t=100\) lattice at \(\kappa=9\) and \(\lambda=1\). (Points are numerical results with tiny error bars connected to guide the eye.)}
\label{fig_partnr}
\end{figure}

In (Fig.~\ref{fig_partnr}) we show a selected result for the particle number density as a function of the chemical potential at zero temperature (i.e., \(1/N_t\) much smaller than other scales involved). The eye-catching feature of this result is the strict independence of \(\mu\), up to some critical value \(\mu_\text{crit}\). This behavior -- also known from other field theories, QCD amongst them -- is known as the Silver Blaze phenomenon \cite{C03}. It is well known (and was checked also in this study c.f.~Section \ref{sec_spec}), that the onset of particle (or anti-particle) surplus takes place at a value of the chemical potential equal to the smallest mass (\(m_{\text{ren}}\)) in the spectrum. A detailed analysis shows us, that this is a second order phase transition at \(\mu_\text{crit} = m_\text{ren} = 1.146(1)\) in lattice units.

\section{2-Point Functions and Spectroscopy at Non-vanishing Chemical Potential}
\label{sec_spec}

The conventional lattice two point correlator
\begin{equation}
\langle \phi_a ~ \phi_b^* \rangle \, = \, \frac{1}{Z} \int \! \mathcal{D}[\phi] ~ e^{-S} ~ \phi_a ~ \phi_b^* \, \equiv \, \frac{1}{Z} ~ Z_{a,b}
\end{equation}
cannot be expressed in terms of partial derivatives of the partition function. Therefore a dual representation of the partition function with field insertions \(Z_{a,b}\) was obtained in \cite{GK13b}. It is very much reminiscent of the partition function \(Z\), with an exception for the arguments of the Kronecker deltas and weight factors at lattice sites \(a\) and \(b\).
\begin{equation}
Z_{a,b} = \sum_{\{k,m\}} ~ \prod_x \delta \left( \sum_\nu \Big[ \ldots \Big]-\,\delta_{x,a}+\delta_{x,b}\right) ~\mathcal{W}\left( \sum_\nu \Big[ \ldots \Big] +\,\delta_{x,a}+\delta_{x,b} \right)~ \Bigg( \ldots \Bigg)~,
\end{equation}
where parts identical to \(Z\) are omitted. The modified constraint gives rise to a new interpretation of the admissible configurations. While in the ensemble \(Z\) only closed loops appear, we now get in addition a single open line starting at lattice point \(a\) and ending at \(b\). \(a\) and \(b\) therefore can be seen as source and sink of \(k\)-flux respectively. In order to sample such configurations we follow \cite{KVW11} and enlarge the ensemble
\begin{equation}
\mathcal{Z} \equiv \sum_{a,b}~Z_{a,b}~,
\end{equation}
which now consists of (all possible) closed loops and open lines. Expectation values with respect to this ensemble are denoted by \(\langle\cdots\rangle_{\mathcal{Z}}\). A slightly modified version of the worm algorithm used for \(Z\) is also applicable for \(\mathcal{Z}\).
The dual expression for the correlator assumes the form
\begin{equation}
\langle \phi_x ~ \phi_y^* \rangle ~ = ~ \frac{Z_{x,y}}{Z} ~ = ~ \frac{\langle \delta_{a,x} ~ \delta_{b,y} \rangle_{\mathcal{Z}}}{\langle \delta_{a,b} ~ \mathcal{W}(n_a)/\mathcal{W}(n_a+2)\rangle_\mathcal{Z}}
\end{equation}
and projected to zero spatial momentum and normalized to \(1\) we find
\begin{equation}
C(t)/C(0) ~ \equiv ~ \frac{\langle \delta_{t,a_4-b_4} \rangle_{\mathcal{Z}}}{\langle \delta_{a_4,b_4} \rangle_{\mathcal{Z}}}~.
\end{equation}

\begin{figure}[h] 
\centering
\includegraphics[width=13cm,type=pdf,ext=.pdf,read=.pdf]{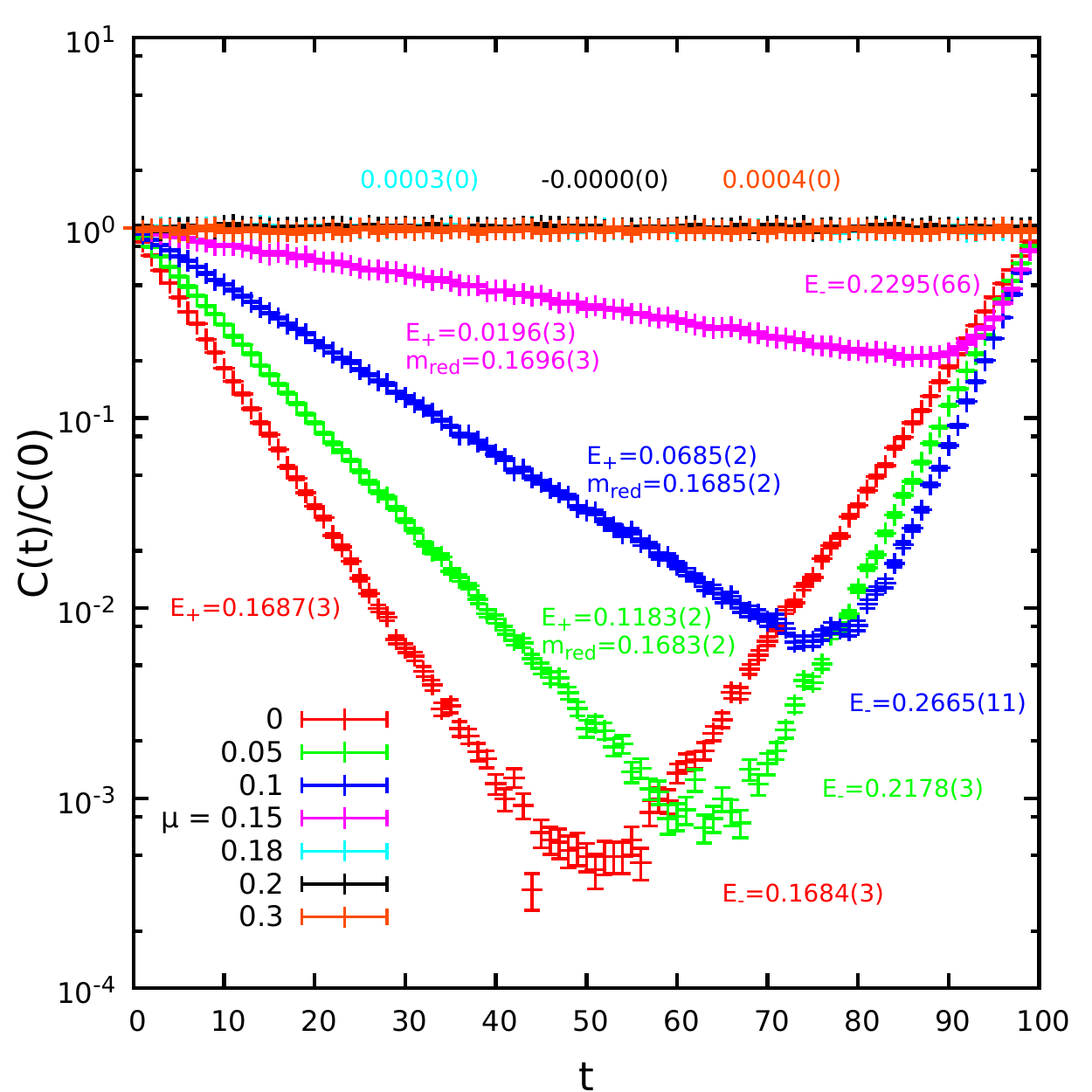}
\caption{Normalized temporal correlator versus time on a \(N_s=32\), \(N_t=100\) lattice at \(\kappa=7.44\) and \(\lambda=1\) for different chemical potentials.}
\label{fig_corr}
\end{figure}

The main result obtained with this method is displayed in (Fig. \ref{fig_corr}) and shows the normalized temporal correlator versus time. The vertical axis has a logarithmic scale and we find straight lines, i.e., exponential decay. The exponents (slopes) for forward and backward propagation are denoted by \(E_{+}\) and \(E_{-}\) respectively. At zero chemical potential we see a symmetric correlator corresponding to \(E_{+} = E_{-}\) within error bars. The situation is qualitatively different at non-zero values of \(\mu\), where the correlator is asymmetric. We find a decreasing slope for forward propagation and an increasing slope for backward propagation. This behavior is expected from the chemical potential -- enhancement of particle propagation and suppression of anti-particle propagation. Both slopes can be related with the help of the reduced mass
\begin{equation}
m_\text{red} = E_{\pm} \pm \mu~,
\end{equation}
which is invariant, as can be seen from the plot and, for example, determined at \(\mu=0\) where \(E_{+} = E_{-} = m_\text{red} = \mu_\text{crit}\) holds. The critical chemical potential (c.f.~Section~\ref{sec_dual}) for this parameter set was found to be \(\mu_\text{crit}=0.170(1)\). For \(\mu\) exceeding this value, the correlator becomes a constant. This is true for the correlator of the field and the conjugate field. A separation of Goldstone and massive modes was not part of this study -- see also the discussion in \cite{GK13b}.

\section{Conclusions, Outlook and Non-Abelian Theories}

It is now established that for the scalar field, \(O(N)\) and \(SU(N)\) symmetric vector models and essentially all models with abelian degrees of freedom one can construct a dual representation in which the complex action problem is absent \cite{DEG11,DEG12,DGS13,DGS13b,GS12,GK13,GK13b}. Fermionic systems with specific interactions can also be mapped to a new representation where the sign problem is solved (e.g., \cite{C12}).

The status of dual representations for models with non-abelian degrees of freedom is, however, unsatisfactory. Previous attempts based on character expansions (e.g. \cite{DZ83}) failed to deliver a positive weight for all configurations. In a recent project we reformulated an \(SU(2)\) spin system with external field, i.e., a system with non-abelian \(SU(2)\) matrix degrees of freedom and compared it to conventional simulations with very good agreement. However, extensions to an \(SU(2)\) gauge model or to the \(SU(3)\) group are missing up to now.

\section*{Acknowledgements}
We thank Y. Delgado Mercado, C. B. Lang and A. Schmidt for discussions. The speaker also wants to thank H.-P. Schadler. This work is partly supported by DFG TR55, \textit{"Hadron Properties from Lattice QCD"} and by the Austrian Science Fund FWF Grant. Nr. I 1452-N27. T. K. is supported by the FWF DK W1203, \textit{"Hadrons in Vacuum, Nuclei and Stars"}.

\end{document}